\documentclass[dvips]{article} 
\usepackage{spie}
\usepackage[dvips]{graphicx}

\newbox\grsign \setbox\grsign=\hbox{$>$} \newdimen\grdimen \grdimen=\ht\grsign
\newbox\simlessbox \newbox\simgreatbox \newbox\simpropbox
\setbox\simgreatbox=\hbox{\raise.5ex\hbox{$>$}\llap
     {\lower.5ex\hbox{$\sim$}}}\ht1=\grdimen\dp1=0pt
\setbox\simlessbox=\hbox{\raise.5ex\hbox{$<$}\llap
     {\lower.5ex\hbox{$\sim$}}}\ht2=\grdimen\dp2=0pt
\def\simgreat{\mathrel{\copy\simgreatbox}}

\title{Sherpa: a mission-independent data analysis application} 

\author{Peter E.~Freeman, Stephen Doe, Aneta Siemiginowska
\skiplinehalf
Harvard-Smithsonian Center for Astrophysics
}

\authorinfo{
Send correspondence 
to: (pfreeman,sdoe,asiemiginowska)@head-cfa.harvard.edu;
phone 1 617 496-7824; fax 1 617 495-7356; Harvard-Smithsonian Center for
Astrophysics, MS 81, 60 Garden St., Cambridge, MA, 02138.
Copyright 2001 Society of Photo-Optical Instrumentation
Engineers.  This paper will be published in Proceedings of SPIE
Conference 4477 and is made available as an electronic preprint with
permission of SPIE.  One print or electronic copy may be made for personal
use only.  Systematic or multiple reproduction, distribution to 
multiple locations via electronic or other means, duplication of any 
material in this paper for a fee or for commercial purposes, or 
modification of the content of the paper are prohibited.}
 
\begin{document} 
  \maketitle 

\begin{abstract}

The ever-increasing quality and complexity of astronomical data
underscores the need for new and powerful data analysis applications.
This need has led to the development of {\it Sherpa}, a
modeling and fitting program in the {\it CIAO} software package that
enables the analysis of multi-dimensional, multi-wavelength data.
In this paper, we present an overview of {\it Sherpa}'s features, which
include:
support for a wide variety of input and output data formats, including
the new Model Descriptor List (MDL) format;
a model language which permits the construction of arbitrarily complex
model expressions, including ones representing instrument characteristics;
a wide variety of fit statistics and methods of optimization, model
comparison, and parameter estimation; 
multi-dimensional visualization,
provided by {\it ChIPS}; 
and new interactive analysis capabilities provided by embedding the
S--Lang interpreted scripting language.
We conclude by showing example {\it Sherpa} analysis sessions.
\end{abstract}



\section{INTRODUCTION}
\label{sect:intro}  

In astronomical data analysis, one develops models of
physical processes in the spectral, spatial, and/or temporal domains, then
fits these models to observed data.  These data may
be of arbitrary dimension, and/or they may
have been collected using an arbitrary number of telescopes that observe
in different wavelength bands.
{\it Sherpa}, the modeling and fitting application of the
{\it Chandra Interactive Analysis of Observations (CIAO)} 
software package,\cite{CIAO} 
is designed to tackle such complex multi-dimensional, 
multi-wavelength analyses.
Free of hard-wired instrument details, 
{\it Sherpa} is outfitted with
a wide variety of models, fit statistics, and methods of optimization,
model comparison, and parameter estimation, and it
offers powerful embedded
visualization, scripting, and data manipulation capabilities.
It can thus be used to analyze energy- or wavelength-space data
from, {\it e.g.}, ground-based telescopes, {\it ISO}, 
{\it Hubble}, {\it XMM}, and the {\it Chandra X-ray Observatory},
as well as from next-generation projects such as
the Virtual Observatory.

In this paper we present a basic overview of {\it Sherpa}.
In {\S}\ref{sect:code}, we describe
the application itself, while in {\S}\ref{sect:func} we describe
{\it Sherpa}'s capabilities using a typical {\it Sherpa} session
as a framework, from reading in data to 
determining the 1$\sigma$ errors on the best-fit model parameters.
In {\S}\ref{sect:examples} we provide brief examples 
of how one may use {\it Sherpa}.
For more information, 
the reader may consult the on-line {\it Sherpa} manual.\cite{ShMan}

\section{IMPLEMENTATION OF SHERPA}
\label{sect:code}

The authors have developed {\it Sherpa} using the
object-oriented C++ programming language.\cite{Doe98}  In object-oriented
programming, objects encapsulate related data and functions; thus,
classes (from which objects are instantiated) can be written that more
closely model the problem domain.  For a modeling and
fitting application, classes containing fitting methods, statistics,
models, and data organize the code in a way that more intuitively
mirrors the concepts underlying the fitting process.

{\it Sherpa} is an interactive application, using a lex/yacc-based parser to
interpret commands.  {\it Sherpa} accepts input via the command-line
interface or ASCII script files.\footnote{
Application developers may also use {\it Sherpa}'s ``wrapper" class,
which provides an API for both C and C++ code.}

The {\it CIAO} package also contains a number of other programs and
libraries that enhance {\it Sherpa}'s capabilities.  Data I/O is provided by
the {\it CIAO} Data Model library.\cite{DM}
The Data Model gives a higher-level,
abstract view of astronomical data files and provides transparent
access to the most common astronomical file formats (FITS, {\it IRAF} IMH,
{\it IRAF} QPOE).\footnote{
ASCII I/O is not provided by the Data Model but is provided
elsewhere in {\it CIAO} library code.}
The Data Model provides a uniform interface so that the
application need no longer use I/O functions specific to particular file
formats.  The Data Model also has a sophisticated filtering and
binning syntax that allows the extraction of selected portions of data
contained in a file.

For all visualization needs, {\it Sherpa} uses 
{\it ChIPS}, the {\it Chandra Imaging
and Plotting System}.\cite{Chips}
As the name indicates, {\it ChIPS} provides an
interface to plotting and imaging applications
(currently {\it SM},\cite{SM} {\it SAODS9},\cite{ds9} and 
{\it SAOTNG}\,\cite{ds9}).
{\it ChIPS} is both a
stand-alone, interactive application and a C++ library (which is
how it is used by {\it Sherpa}--it
passes commands and data to {\it ChIPS}).
{\it Sherpa} can access {\it ChIPS} functions through
the {\it ChIPS} API, but it is also possible for {\it ChIPS} commands to be 
input directly
at the {\it Sherpa} command-line.

As of {\it CIAO 2.1}, the S--Lang interpreted scripting language\cite{Slang}
has been embedded in {\it CIAO}.  Its features include:
\begin{itemize}
\itemsep=-4pt
\item{global and local interpreted variables, and multidimensional arrays
(up to seven dimensions);}
\item{branching and looping, and programmability with user-defined functions;}
\item{string and numeric data types, structures, and a limited form of
pointers;}
\item{built-in arithmetic and mathematical functions, which operate
transparently on arrays; and}
\item{extensibility--the ability to create new functionality for {\it CIAO}
applications ({\it e.g.} GUIDE; see {\S}\ref{sect:chgrating}).}
\end{itemize}
S--Lang is accessed through a supplementary library layer, dubbed
Variables, Math, and Macros (VARMM),\cite{Doe01} which gives the
user additional capability to define structured variables directly
from disk files, as well as enabling existing {\it CIAO} applications
to access S--Lang variables and other features with a bare minimum of
new development effort.

\section{SHERPA FUNCTIONALITY} 
\label{sect:func}

The capabilities of {\it Sherpa} can be best described by following the
steps of a typical analysis session.
In such a session, the user would:
\begin{itemize}
\itemsep=-4pt
\item{read in (source and/or background) data (and set filters, {\it etc.});}
\item{build models that describe these data (as well as the telescope/detector
combination);}
\item{choose a statistic that quantifies how well these
models describe the data;}
\item{fit the models to the data, determining the
minimum value of the chosen statistic;}
\item{compare best-fit results achieved with different models to
select one best-fit model; and}
\item{estimate the errors for each parameter of the best-fit model.}
\end{itemize}
Below, we discuss each of these items in turn.  We note that a 
typical session would also include {\it ChIPS}
visualization (of the source data,
or of the background fit, {\it etc.});
while we do not discuss it specifically in this section,
we provide examples of visualization in {\S}\ref{sect:examples}.

\subsection{Sherpa Data Input (and Output)} 
\label{sect:io}

Data input marks the beginning of a {\it Sherpa} analysis session.
The data may be input from either a file on disk or an interpreted S--Lang
variable.  In Table \ref{tab:io}, we list data types
that may be input into {\it Sherpa},
while in Table \ref{tab:iotypes} we list currently supported file formats.
(Instrument characteristics, such as the point-spread function, or PSF,
may also be contained in files that are
read in when an instrument model is specified; see {\S}\ref{sect:inst}.)

\begin{table} [ht]   
\caption{Input data types.}
\label{tab:io}
\begin{center}
\begin{tabular}{|l|l|} 
\hline
\rule[-1ex]{0pt}{3.5ex} {\tt DATA} & Source data \\
\hline
\rule[-1ex]{0pt}{3.5ex} {\tt BACK} & Background data \\
\hline
\rule[-1ex]{0pt}{3.5ex} {\tt (B)ERRORS} & Source or background errors \\
\hline
\rule[-1ex]{0pt}{3.5ex} {\tt (B)ERRORS SYSTEM} & Source or background systematic errors \\
\hline
\rule[-1ex]{0pt}{3.5ex} {\tt FILTER} & Specification of which bins of a given dataset are to be analyzed \\
\hline
\rule[-1ex]{0pt}{3.5ex} {\tt WEIGHT} & Statistical weights for each datum \\
\hline
\rule[-1ex]{0pt}{3.5ex} {\tt GROUPS} & Specification of how data are to be binned \\
\hline
\end{tabular}
\end{center}
\end{table}

\begin{table} [ht]   
\caption{Supported file types.}
\label{tab:iotypes}
\begin{center}
\begin{tabular}{|l|l|} 
\hline
\rule[-1ex]{0pt}{3.5ex} {\tt ASCII} & ASCII data \\
\hline
\rule[-1ex]{0pt}{3.5ex} {\tt FITSBIN} & FITS binary table \\
\hline
\rule[-1ex]{0pt}{3.5ex} {\tt FITS} & FITS image \\
\hline
\rule[-1ex]{0pt}{3.5ex} {\tt IMH} & {\it IRAF} IMH image \\
\hline
\rule[-1ex]{0pt}{3.5ex} {\tt PHA} & Type I \& II PHA data \\
\hline
\rule[-1ex]{0pt}{3.5ex} {\tt QP} & {\it IRAF} QPOE image \\
\hline
\end{tabular}
\end{center}
\end{table}

One only needs to read in source data to start an analysis session;
other data types (such as background) are not required.
However, some types of
data may be input automatically when source data are read; for
instance, a PHA file can have columns that
specify the statistical and systematic errors ({\tt STAT\_ERR} and
{\tt SYS\_ERR}, respectively) and the data grouping ({\tt GROUPING}),
and it can have a keyword ({\tt BACKFILE}) specifying the
background dataset.
Also, one can specify statistical and systematic errors, filters, and 
statistical weights via the command-line interface.
If statistical errors are not input or specified, they are
estimated by {\it Sherpa} during fitting; see {\S}\ref{sect:stat}.

An arbitrary number of datasets may be input into {\it Sherpa}, and
arbitrary subsets of these data may be jointly analyzed.  Since current
standard processing of {\it Chandra} grating data includes the
extraction of background spectra from regions on either side of
the source extraction region,
one may also specify up to two background datasets per source dataset.  
These data may be fit
simultaneously with the source data (see {\S}\ref{sect:srcback}), or 
they may be subtracted from the source data on a channel-by-channel basis 
prior to a fit:
\begin{equation}
S_i'~=~S_i - \beta_S t_S \left[ \frac{\sum_{j=1}^N B_{i,j}}{\sum_{j=1}^N \beta_{B_j} t_{B_j}} \right] \,.
\end{equation}
$S_i$ is the source datum in bin $i$, 
$B_{i,j}$ is the background datum in bin $i$ of background set $j$,
$t$ is the observation time, and $\beta$ is the ``backscale" 
(the {\tt BACKSCAL} header keyword value in a PHA file),
typically
defined as the ratio of data extraction area to total detector area.



At any time during an analysis session, quantities like the
background-subtracted data, convolved model
amplitudes, or fit residuals may be output to files
using the {\tt WRITE} command.  (The files may be saved in any of the
formats listed
in Table \ref{tab:iotypes}.)
A user may also save (and later restore) the state of a
{\it Sherpa} analysis session using a Model Descriptor List (MDL).
The MDL is a record of all
information relevant to a Sherpa fitting session: the names of all
input data files and associated filters; all defined
source and instrument models, with model parameter settings;
the choices of optimization method and fit statistic;
and the fluxes and identification of emission/absorption
lines that the user may have identified (see {\S}\ref{sect:chgrating}).
The MDL may be either saved to disk or instantiated as an
S--Lang variable.

\subsection{Building Model Expressions} 
\label{sect:model}

Once source and/or background
data are input, the next step is to create
model expressions reflecting one's knowledge of the physical
processes which gave rise to those data.  (The user can also
build model expressions that represent the observing instruments.
See {\S}\ref{sect:inst}.)
{\it Sherpa} currently provides nearly 40 of its own
one- and two-dimensional models
and 90 one-dimensional
{\it XSPEC}\,\cite{Arn96} models that may be arbitrarily
combined to build complex composite models, as we show below.
Note that a user can also define a model\footnote{
The {\tt usermodel}.  One can also define
an optimization method ({\tt usermethod}) and
statistic ({\tt userstatistic}).} and compile it
into the {\tt libascfitUser.so} shared-object library,\footnote{
In {\it CIAO} 2.2, {\it Sherpa} will also
support user-model definition via S--Lang scripts.}
where it can be accessed by {\it Sherpa}.

\newpage

\paragraph{Sherpa model language.}
The {\it Sherpa} model language resolves ambiguity by allowing the user
to give a unique name or alias to each instance of a model.  For
example, if two datasets are entered, and each is to be fit with
a Gaussian model, but with different parameters, one might type:\footnote{
The relationship between {\tt gauss1d}, and {\tt g1} and {\tt g2} above
is similar to the class-object relationship in object-oriented 
programming: {\tt gauss1d} is the class, specifying a Gaussian
function with parameters position, amplitude, and full-width at half-maximum,
while {\tt g1} and {\tt g2} are instantiated objects of the class, each
containing a specific set of parameter values.}
\begin{verbatim}
sherpa> gauss1d[g1]
sherpa> gauss1d[g2]
sherpa> source 1 = g1
sherpa> source 2 = g2
\end{verbatim}
whereas if each is to be fit with the same Gaussian model, one might type:
\begin{verbatim}
sherpa> gauss1d[g1]
sherpa> source 1:2 = g1
\end{verbatim}

\paragraph{Linking parameters to other parameters or to models.}
One can link an individual model parameter to another model parameter, 
so that their values are correlated.
For instance, if a particular atomic line is observed by
two different detectors, it could be modeled with two Gaussian functions
whose centroids are linked:
\begin{verbatim}
sherpa> source 1 = gauss1d[g1]
sherpa> source 2 = gauss1d[g2]
sherpa> g1.pos => g2.pos
\end{verbatim}
(Note that simple arithmetic relations are also possible, 
{\it e.g.}~{\tt g1.ampl => 2*g2.ampl}.)
At this point, {\tt g1.pos} is no longer a free parameter of the fit.
It can be made free again with the 
command {\tt UNLINK g1.pos}.

One can also link an individual model parameter to a model,
to describe how a parameter's value will vary as a function of position
in parameter space.  For 
instance, one can model emission from an accretion disk using a blackbody
function whose temperature is a function of radius:
\begin{verbatim}
sherpa> source = bb
sherpa> Temperature = POLY
sherpa> bb.kT => Temperature
\end{verbatim}

\paragraph{Nesting model expressions.}
A model may be nested within another, {\it i.e.} one may specify a
model expression of the form g(f(x)).
In this example, the
input data axis is transformed to log-space using {\it Sherpa}'s
log model, and a blackbody model is evaluated in that space:
\begin{verbatim}
sherpa> logenergy = shlog
sherpa> source = bb{logenergy}
\end{verbatim}

\paragraph{Multi-dimensional model expressions.}
One can specify completely different models that 
are to be evaluated along different
axes of a multi-dimensional dataset, as in this example, where two-dimensional
spectral-radius data are modeled with a combination of Lorentzian
and power-law models:
\begin{verbatim}
sherpa> lorentz[Spatial]
sherpa> pow[Spec]
sherpa> source = Spatial{x1}*Spec{x2}
\end{verbatim}
{\tt x1} and {\tt x2} represent the first and second axes of the
input image, respectively.

\subsubsection{Instrument models}
\label{sect:inst}

Instrument models are used to quantify 
characteristics, such as effective
area, a detector's energy response, or a mirror's point-spread function. 
They provide a mapping from photon space (where source and background
models are evaluated) to counts space (where fit statistics
are computed).  The instrument model class is the key element which
makes {\it Sherpa} a mission-independent application, permitting analysis
of data observed by any telescope, regardless of whether it is ground-based
or space-based.

Currently, {\it Sherpa} defines three instrument model classes:
(1) {\tt RSP}, in which an evaluated one-dimensional model is multiplied
by an ancillary response (ARF, {\it i.e.}~an effective area) on a bin-by-bin
basis, then folded through a response matrix (RMF); (2) {\tt PSFFromFile}, in 
which the evaluated one- or two-dimensional model is convolved with a
numeric kernel; and (3) {\tt PSFFromTCD}, in which the evaluated 
one- or two-dimensional model is convolved with an analytic kernel
({\it e.g.}~Gaussian) defined within {\it CIAO}'s Transformation,
Convolution, and Deconvolution (TCD) library.  Future versions of
{\it Sherpa} may include new classes to treat, {\it e.g.}, two-dimensional
exposure maps.

\subsection{Statistics} 
\label{sect:stat}

\subsubsection{Statistics based on the $\chi^2$ distribution}

The $\chi^2$ statistic is appropriate for the analysis of 
Gaussian-distributed data.  It is defined as
\begin{eqnarray*}
\chi^2~&\equiv&~\sum_i \frac{(D_i - M_i)^2}{\sigma_i^2} \,.
\end{eqnarray*}
where $D_i$ is the (source or background) data in bin $i$,
$M_i$ is the (convolved source or background) model 
predicted amplitude in bin $i$,
and $\sigma_i$ is the estimated error for the $i^{\rm th}$ datum
(the square root of the variance of the distribution from
which that datum had been sampled).
As noted in {\S}\ref{sect:io}, one may specify the errors via
a file or the command-line; if this
is done, the $\chi^2$ statistic is used as shown above.  Otherwise, 
the data are assumed to be Poisson-distributed,\footnote{
The Poisson distribution tends asymptotically towards a
Gaussian distribution as its expectation value approaches infinity.}
with the errors for each datum estimated during analysis.
The large array of error estimators that {\it Sherpa} provides
is one of its key features; these are listed in Table \ref{tab:stat}.
Note that the entries in this table are only correct if the background data
have not been subtracted from the source data; otherwise errors are propagated
in the standard manner ($\sigma_{S'}^2 = \sigma_S^2 + \sigma_B^2$).
Also note that error estimates based on model amplitudes are inappropriate to
use in the analysis of background-subtracted data, as they generally
underestimate the true error.

Using a $\chi^2$-based statistic to analyze counts
data is generally only valid in the Gaussian (high-counts) limit
($\simgreat$ 5 counts in {\it each} bin).
This is because the approximations that must be made to derive the
$\chi^2$ statistic from Poisson log-likelihood
${\log}{\cal{L}}$ break down otherwise.
The {\tt CHI GEHRELS}\cite{Geh86} and {\tt CHI PRIMINI}\cite{KPA95} 
statistics are designed to work with low-count data; note that the 
former it is not generally sampled from the $\chi^2$ distribution
and thus the derived best-fit statistic may appear to be ``too good"
($\chi_G^2 \slash N \ll$ 1, where $N$ is the number of degrees of
freedom in the fit), in the low-counts limit.

\subsubsection{Statistics based on the Poisson likelihood}

The Poisson likelihood function is
\begin{equation}
{\cal{L}}~=~\prod_i \frac{M_i^{D_i}}{D_i!}\exp(-M_i) \,.
\end{equation}
{\it Sherpa} features two statistics based on this function:
{\tt CASH} and {\tt BAYES}.  

The version of the {\tt CASH} statistic\cite{Cash79} used by {\it Sherpa}
is derived from $\cal{L}$ by (1) taking its logarithm,
(2) dropping the factorial term (which remains
constant during fits to given datasets), (3) multiplying
by two, and (4) changing the sign (so that the statistic may be
minimized, like $\chi^2$):
\begin{equation}
C~\equiv~2 \sum_i [ M_i - D_i \log M_i ] \,,
\end{equation}
In the high-counts limit, ${\Delta}C \sim {\Delta}\chi^2$, so that
in principle one can use ${\Delta}C$ instead of ${\Delta}\chi^2$ in
model comparison tests (see {\S}\ref{sect:modcomp}).

\begin{table} [ht]   
\caption{Statistics based on the $\chi^2$ distribution.}
\label{tab:stat}
\begin{center}
\begin{tabular}{|l|l|} 
\hline
\rule[-1ex]{0pt}{3.5ex} Statistic & Variance $\sigma_i^2$ \\
\hline
\hline
\rule[-1ex]{0pt}{3.5ex} {\tt CHI DVAR} & $D_i$ \\
\hline
\rule[-1ex]{0pt}{3.5ex} {\tt CHI GEHRELS} ({\it Sherpa} default) & $\left[ 1 + \sqrt{D_i+0.75} \right]^2$ \\
\hline
\rule[-1ex]{0pt}{3.5ex} {\tt CHI MVAR} & $M_i$ \\
\hline
\rule[-1ex]{0pt}{3.5ex} {\tt CHI PARENT} & $(\sum_{i=1}^N D_i) \slash N$ \\
\hline
\rule[-1ex]{0pt}{3.5ex} {\tt CHI PRIMINI} & $M_i$ from previous best-fit \\
\hline
\end{tabular}
\end{center}
\end{table}

The {\tt BAYES} statistic\cite{Lor92} is based on Bayesian statistical 
methodology\footnote{
Space does not permit us to provide details about Bayesian statistical
methodology, which may be less familiar to some readers than the
standard ``frequentist" statistical paradigm.
For an introduction to Bayesian statistics that is geared towards
astrophysicists, see Loredo (1992)\cite{Lor92}.}
and is appropriate to use when a background
is input and the rate of background accumulation may be taken as
the same in both the background and source extraction regions.
This statistic takes into account uncertainty
in the (implicitly defined) background amplitudes via marginalization:
\begin{eqnarray*}
B~\equiv~-p({\vec x}_S \vert D) = -\sum_i \int_{x_{B,i}} dx_{B,i} p({\vec x}_S,x_{B,i} \vert D) \,,
\end{eqnarray*}
where ${\vec x}_S$ represents the set of source model parameters and 
$x_{B,i}$ is the background amplitude in
the $i^{\rm th}$ bin.  (Note that the above equation has an analytic
solution that we do not reproduce here.)

Note that because the {\tt CASH} and {\tt BAYES} statistics are based on the 
likelihood function, they should not be applied to background-subtracted data.
Also, there is no ``goodness-of-fit" measure associated with 
{\tt CASH} and {\tt BAYES}, as there is of $\chi^2$-based statistics.
Such a measure can, in principle, be computed
by performing Monte Carlo simulations: one would repeatedly sample new
datasets from the best-fit model, fit them, and note where the observed
statistic lies within the derived distribution of statistics.

\subsection{Optimization} 
\label{sect:opt}

Optimization is the act of minimizing $\chi^2$ or 
$-{\log}{\cal{L}}$ by varying the thawed parameters of the
defined model.  {\it Sherpa} provides a number of optimization methods,
which can be classified in two broad
categories: those which find a local minimum of the statistical surface in 
parameter space by moving along the local gradient of that surface,
and those which examine large (hyper-)volumes of parameter space
in a search for the global minimum (see
Table \ref{tab:opt}\footnote{
Along with these {\it Sherpa} methods, a future release of {\it CIAO}
will feature a stand-alone
fitting application for low-counts data 
which uses Bayesian posterior sampling.  See
van Dyk et al.~(2001).\cite{vD01}}).

Below, we discuss the three optimization methods appropriate for finding 
local minima: {\tt POWELL}, {\tt SIMPLEX}, and {\tt LEVENBERG-MARQUARDT}.
Users should be acquainted with
the (dis)advantages of each so as to make the best use of them.
(For more information about {\it Sherpa's} other optimization methods, consult
the {\it Sherpa} manual\cite{ShMan} and, {\it e.g.}, Press 
{\it et al.}~1992\cite{Pr92}.)

{\tt POWELL}, a direction-set method in which the chosen statistic is
minimized by varying each member of an (initially orthogonal) set of 
parameter-space vectors in turn, is {\it Sherpa's} default optimizer.
Its advantages include the fact that no gradient calculations are required,
and that it is a robust method, capable of finding minima even on complex
statistical surfaces.  (Also, unlike {\tt LEVENBERG-MARQUARDT}, is can
be used effectively with likelihood-based statistics.)  Its primary
disadvantage is that it is relatively slow.

\begin{table} [ht]   
\caption{Optimization methods in {\it Sherpa}.}
\label{tab:opt}
\begin{center}
\begin{tabular}{|l|l|} 
\hline
\rule[-1ex]{0pt}{3.5ex} Local Minimum & {\tt POWELL}, {\tt SIMPLEX}, {\tt LEVENBERG-MARQUARDT} \\
\hline
\rule[-1ex]{0pt}{3.5ex} Global Minimum & {\tt GRID(-POWELL)}, {\tt MONTE(-POWELL)}, {\tt SIMULATED ANNEALING} \\
\hline
\end{tabular}
\end{center}
\end{table}

In {\tt SIMPLEX} optimization, the fit statistic is calculated at 
the $N+1$ vertices of a simplex in a $N$-dimensional parameter space,
with the vertices being moved until the local minimum is bracketed.
Its advantages include the fact that no gradient calculations are required,
it can find minima of complex statistical surfaces, and it requires fewer
model evaluations than {\tt POWELL}.  However, it is not as
robust as {\tt POWELL}.  The {\tt SIMPLEX} method is best-used when one
starts the optimization close to the local minimum; for instance, it
is a good optimizer to use in parameter estimation (see {\S}\ref{sect:parest}).

In {\tt LEVENBERG-MARQUARDT} optimization, the local minimum is 
approached by taking steps in parameter space 
whose magnitudes ${\delta}{\vec x}$ 
are computed by solving the set of linear equations 
\begin{eqnarray*}
\sum_{j=1}^n \alpha_{ij}(1+\lambda_{ij}){\delta}x_j = \beta_i \,,
\end{eqnarray*}
where
\begin{eqnarray*}
\alpha_{ij}~=~\sum_{k=1}^n \frac{1}{\sigma_k^2} \left[ \frac{{\partial}M({\vec x})}{{\partial}x_i} \frac{{\partial}M({\vec x})}{{\partial}x_j} \right]~~{\rm and}~~\beta_i~=~-\frac{1}{2}\frac{{\partial}\chi^2}{{\partial}x_i} \,,
\end{eqnarray*}
and $\lambda_{ij}$ is a matrix with non-zero diagonal elements whose
magnitudes are inversely proportional to ${\delta}{\vec x}$.
The primary advantage of {\tt LEVENBERG-MARQUARDT} optimization is
speed, while its disadvantages include the fact that a gradient
computation is required, that it is appropriate for use with 
$\chi^2$-based statistics only, and that it is less robust when
applied to optimization on a
complex statistical surface.  (To circumvent the third
issue, we have introduced the option that the optimization method
may be switched from {\tt LEVENBERG-MARQUARDT} to {\tt SIMPLEX} close
to the minimum, where the disadvantages of {\tt LEVENBERG-MARQUARDT} become
more readily apparent.)

\subsection{Model Comparison} 
\label{sect:modcomp}

Often, a user will fit more than one parametrized model to a given dataset,
and will wish to compare the best-fit results of each.  For instance,
one may fit two continuum models to data,
and need to decide whether the improvement in the fit statistic
that is observed when using the more complex model
is attributable to chance.  To make this decision, one uses a
model comparison test to yield either: (1) the frequentist test
significance, $\alpha$, which is the probability of selecting the
alternative (more complex) model $M_1$ when in fact the null hypothesis
$M_0$ is correct; or (2) the Bayesian odds, the ratio of 
model posterior probabilities 
for $M_1$ and $M_0$.
If the prior probability distribution for 
a model's parameter values is constant, then its
posterior probability is proportional to
the integral of the likelihood function 
$\cal{L}$ over parameter space.

The model comparison test that is currently available to the {\it Sherpa}
user is the $\chi^2$ Goodness-of-Fit (GOF) test, an alternative-free
test.  The next version of {\it Sherpa} will also
contain the Maximum Likelihood Ratio (MLR) test and the $F$-test.
Methods of model comparison that may be included in future versions of 
{\it Sherpa} include: using simulations 
to determine model comparison test
statistics numerically when the conditions for using an analytic test
are not fulfilled; computing the Bayesian odds using the
Laplace approximation\cite{LL92}; and computing the Bayesian odds
via numerical integration.  We note that these new techniques, in addition to
assisting the comparison of models, would also be useful for parameter
estimation.

\subsection{Parameter Estimation} 
\label{sect:parest}

Once one has selected a best-fit model, the next question is: what are
the errors on the model parameters, {\it i.e.}~what
are the confidence intervals associated with each model parameter?
In general, a frequentist statistician can determine possible intervals 
by repeatedly simulating
data from the best-fit model, fitting these data, and determining the
distribution of best-fit values for each model parameter.\footnote{
A Bayesian would adapt methods mentioned in the previous section--using
numerical integration or the Laplace approximation, {\it etc.}--to
the problem of parameter estimation.  Thus in the remainder of this 
subsection, we only discuss frequentist parameter estimation methods.}
The central
68\% of each distribution can then be deemed the 1$\sigma$ confidence
interval.  However, simulations are computationally expensive, and
if: (1) the $\chi^2$ or ${\log}{\cal{L}}$ surface is approximately
shaped like a multi-dimensional paraboloid ({\it i.e.}~contours of
constant $\chi^2$ or ${\log}{\cal{L}}$ appear ellipsoidal in 
two-dimensional plots), and (2) the best-fit point
is sufficiently far from parameter space boundaries, then 
confidence intervals may be estimated by examining the statistical
surface itself.  

{\it Sherpa} currently features three parameter estimation methods
appropriate for use when the statistical surface is ``well-behaved":
{\tt UNCERTAINTY}, {\tt PROJECTION}, and {\tt COVARIANCE}.  
(In addition, one can make one- or two-dimensional plots showing
the fit statistic value as a function of parameter value[s].) 
With {\tt UNCERTAINTY}, the error for a particular thawed parameter
is estimated by varying its value (while holding all other parameter
values fixed to their best-fit values)
until the fit statistic increases by a preset amount from its minimum
value ({\it e.g.}~${\Delta}\chi^2$ = 1 for 1$\sigma$).
{\tt PROJECTION} is similar to {\tt UNCERTAINTY}, except that the
values of all other parameters are allowed to float to new best-fit
values.  With {\tt COVARIANCE}, errors are estimated by calculating the
covariance matrix, the inverse of the matrix of statistical surface
second derivatives at the best-fit point.  Each of these methods has
distinct (dis)advantages: for example, {\tt UNCERTAINTY}, while fast,
will generally underestimate an interval's size if the parameter is
correlated with other parameters; and {\tt PROJECTION} provides a means
to visualize the surface and can be used even if the model parameters
are correlated, but is in the strictest statistical sense no
more accurate than the much faster {\tt COVARIANCE} method (which is
itself not useful for visualization).

\section{EXAMPLES OF SHERPA ANALYSES} 
\label{sect:examples}

In this section, we present four examples of {\it Sherpa} analyses.
We note that space limitations prevent us from showing all but a few
commands that are used in these analyses; for full scripts, plus
scripts showing other analyses, please
consult the {\it Sherpa} analysis threads.\cite{ShThr}

\subsection{Multi-wavelength analysis of spectra}
\label{sect:hstsax}

\begin{figure}
\begin{center}
\begin{tabular}{c}
\includegraphics[height=8cm]{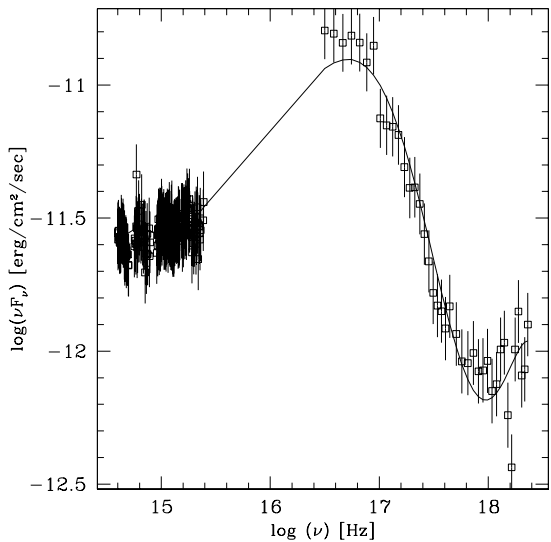}
\end{tabular}
\end{center}
\caption[hstsax]
{\label{fig:hstsax}
Best-fit of two polynomial functions to data of the narrow-line
Seyfert 1 galaxy RE J1034+396, observed by 
the William Herschel 4.2m Telescope, {\it HST}, and {\it BeppoSAX}.}
\end{figure}

In this example, we analyze flux data (log$[{\nu}F_{\nu}]$) of 
RE J1034+396,\cite{P01}
a low-redshift, narrow-line Seyfert 1 galaxy.  The data were collected by
the William Herschel 4.2m Telescope, {\it HST},
and {\it BeppoSAX}.  Because the data are not sampled from a Poisson
distribution, the errors must be input or specified; here, we assume that
the error on the flux is 1\%:
\begin{verbatim}
sherpa> errors = 0.01*data
\end{verbatim}
The observed ``blue bump" is modeled in {\it Sherpa} with
two polynomial functions; the final fit 
is shown in Figure \ref{fig:hstsax}.

\subsection{Simultaneous analysis of source and background data}
\label{sect:srcback}

In this example, we analyze {\it Chandra} source and background spectra
of the supernova remnant G21.5--0.9.  In our analysis, we
assume a power-law times galactic absorption model, with different model
parameters for the source and background:
\begin{verbatim}
sherpa> source = xswabs[sabs]*pow[sp]   # uses the XSPEC wabs absorption model
sherpa> bg = xswabs[babs]*pow[bp]
\end{verbatim}
We model the source and background data separately, rather than subtract the
background data from the source data, because the low background count-rate.
This low count-rate also motivates the use of the Cash statistic:
\begin{verbatim}
sherpa> statistic cash
\end{verbatim}
The final fit is shown below, and in Figure \ref{fig:backg}.
\begin{verbatim}
sherpa> fit
 powll: v1.2
 powll:     converged to minimum =    -7.01375E+03 at iteration =     28
 powll:   final function value    =    -7.01375E+03
          sabs.nH  2.38646  10^22/cm^2  
            sp.gamma  1.50622     
            sp.ampl  0.00201939     
          babs.nH  0.629181  10^22/cm^2  
            bp.gamma  1.0345     
            bp.ampl  0.000101356     
\end{verbatim}

\begin{figure}
\begin{center}
\begin{tabular}{c}
\includegraphics[height=8cm]{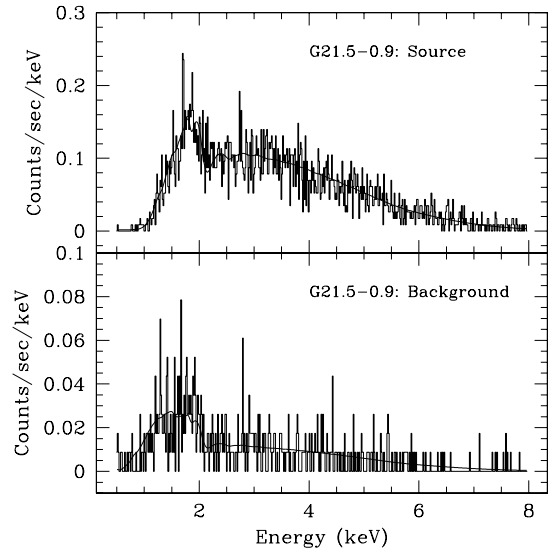}
\end{tabular}
\end{center}
\caption[backg]
{\label{fig:backg}
{\it Top:} Best-fit of a power-law times galactic absorption model
to the source spectrum of supernova remnant G21.5--0.9.  {\it Bottom:}
Best-fit of a different power-law times galactic absorption model
fit to a background spectrum extracted near G21.5--0.9.
}
\end{figure}

\subsection{Analysis of Chandra grating data}
\label{sect:chgrating}

\begin{figure}
\begin{center}
\begin{tabular}{c}
\includegraphics[height=8cm]{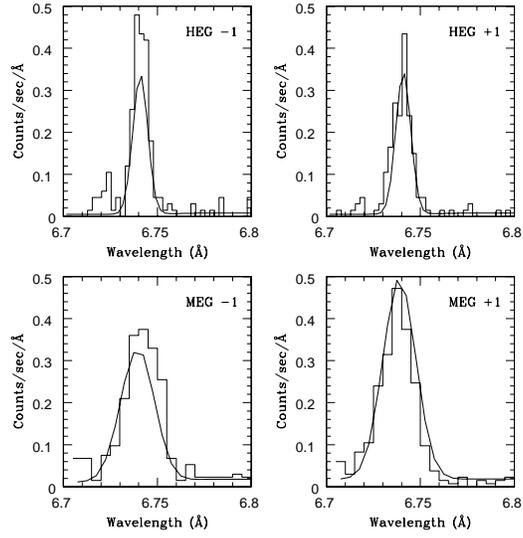}
\end{tabular}
\end{center}
\caption[chgrating]
{\label{fig:chgrating}
Best-fit of a normalized Gaussian function to an emission line
(Si XIII 2$\rightarrow$1 at 6.7403 \AA) observed in four first-order
HEG and MEG {\it Chandra} grating spectra of Capella.}
\end{figure}

This example shows the analysis of
{\it Chandra} grating spectra of the bright
X-ray source Capella, which have been stored in one Type II PHA file.
We concentrate on the first-order
High Energy Grating (HEG) and Medium Energy Grating (MEG) spectra,
which are input into {\it Sherpa} as datasets 3 (HEG -1), 4 (HEG +1),
9 (MEG -1), and 10 (MEG +1).
Because the input Type II PHA data file contains columns defining the
wavelengths for each bin, the analysis is assumed to be in wavelength-space.
We examine only data between 6.7 and 6.8~\AA:
\begin{verbatim}
sherpa> notice allsets wave 6.7:6.8
\end{verbatim}
We then fit a normalized Gaussian function to the observed line:
\begin{verbatim}
sherpa> source 3,4  = ngauss[hg1] + const[co]
sherpa> source 9,10 = ngauss[mg1] + co
\end{verbatim}
where the constant function represents the background.
Because the line flux will be same in a contemporaneous MEG/HEG
observation, the amplitudes are linked:
\begin{verbatim}
sherpa> mg1.ampl => hg1.ampl
\end{verbatim}
Other parameters are not linked because of uncertainties in
calibration.  Note that we use only grating ARFs in this analysis;
we could also model the line profile with a delta function and
use both grating ARFs and RMFs.
After the fit (Figure \ref{fig:chgrating}), 
we identify the most likely transition which gives rise
to the observed line using
GUIDE, a S--Lang-based extension to {\it Sherpa} which acts as an
interface to the Atomic Plasma Emission Database (APED):\cite{APED}
\begin{verbatim}
sherpa> import("guide")
sherpa> identify(6.40)
Found 9 lines.
  Lambda  --    Ion     UpperLev  LowerLev Emis(ph cm^3/s) @ Peak Temp
   ...
   6.7403 --    Si XIII      2 ->      1,  3.548e-17 @ logT =  7.00
   ...
\end{verbatim}

\subsection{Analysis of two-dimensional data}
\label{sect:twod}

\begin{figure}
\begin{center}
\begin{tabular}{c}
\includegraphics[height=10cm]{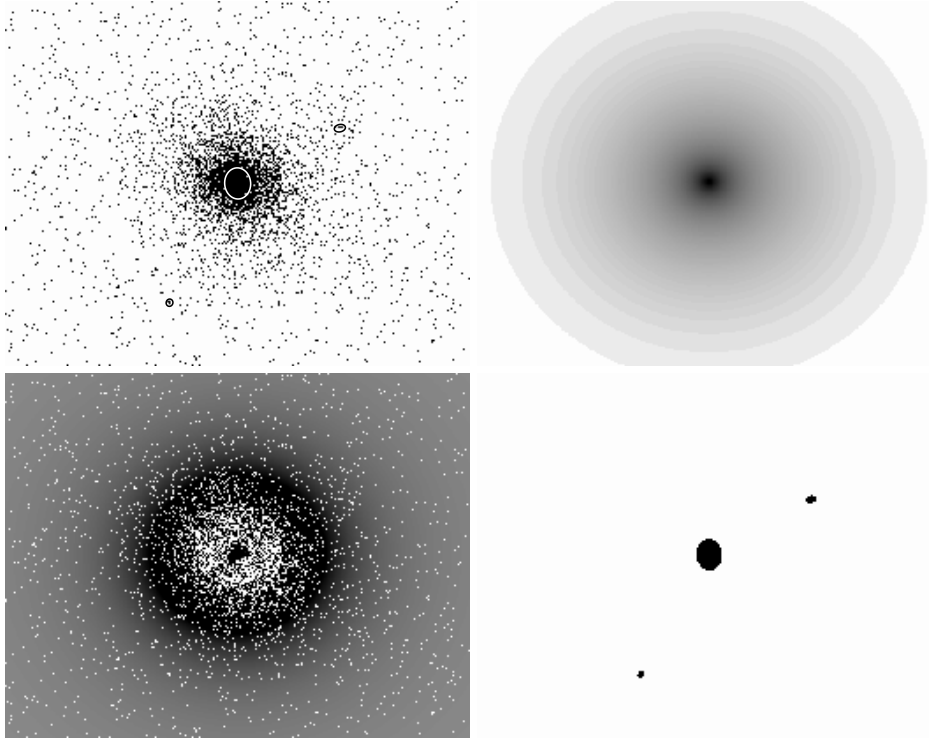}
\end{tabular}
\end{center}
\caption[ch2d]
{\label{fig:ch2d}
{\it Top Left:} {\it Chandra} ACIS-S data of X-ray cluster
MS 2137.3-2353, with {\it SAODS9} source regions
superimposed.  {\it Top Right:} Best-fit of a two-dimensional
beta model to the filtered data.  {\it Bottom Left:}
Residuals (in units of $\sigma$) of the best fit.
{\it Bottom Right:} The applied filter; the data within
the ovals were excluded from the fit.
}
\end{figure}

In this last example, we demonstrate how one can model the
spatial distribution of hot gas in the X-ray cluster
MS 2137.3-2353, observed by {\it Chandra}.
After the data are entered, we display them
using {\it SAODS9}\,; we then load the three point
source regions 
into {\it SAODS9} that we will use to interactively filter the data:
\begin{verbatim}
sherpa> ignore image
\end{verbatim}
One could also use the regions to filter the data directly at the
{\it Sherpa} command line:
\begin{verbatim}
sherpa> ignore filter ellipse(300.14946,299.8716,20.128119,16.76774,94.648547) + \
                      ellipse(431.96938,371.1944, 7.251325, 4.77655,11.890284) + \
                      ellipse(212.26666,145.8972, 4.744431, 4.33702,71.631001) 
\end{verbatim}
This command filters out the data {\it within}
the defined regions.  The remaining data, which represent only the
intra-cluster gas, are then fit with a 
two-dimensional beta model profile.  See Figure \ref{fig:ch2d}.

\acknowledgments     
 
We would like to thank Mark Birkinshaw, 
William Joye, Malin Ljungberg, and Michael Noble for past and
present contributions to {\it Sherpa}'s development.
We would also like to thank Holly Jessop for her tireless work
maintaining the {\it Sherpa} manuals and threads.
The {\it Sherpa} project is supported by the {\it Chandra} X-ray
Center under NASA contract NAS8-39073.



\end{document}